\documentclass[12pt,a4paper]{article}
\newcommand{\del}{\partial}
\newcommand{\mbfit}[1] {\mbox{\boldmath{$#1$}}}
\newcommand{\schreq}{Schr\"{o}dinger equation}
\newcommand{\half}{\frac{1}{2}}
\newcommand{\NABLA}{\mbfit{\nabla}}
\newcommand{\relval}[1] {\ensuremath{#1_\mathrm{rel}}}

\begin{document}  
%\newcounter{num}
%\newcounter{num1}
              
\begin{center}                  
{\large\bf Improving our understanding of the Klein--Gordon equation} \vspace*{3.0mm}\\ 
P. J. Bussey.\\
School of Physics and Astronomy, University of Glasgow, Glasgow G12 8QQ, U.K.
\end{center}
E-mail: peter.bussey@glasgow.ac.uk\\
Orcid: 0000-0003-0988-7878\\[3mm]
%The author has no conflicts of interest.

\noindent
{\bf Abstract}\\ A detailed consideration of the Klein-Gordon equation 
in relativistic quantum mechanics
is presented in order to offer more clarity than many standard
approaches. The equation is frequently employed in the research literature, 
even though  problems have  
often been raised regarding its second-order nature, the status of its 
negative-energy solutions and the formulation of particle density and
flux.  Most of these problems can be avoided by dismissing the
negative-energy solutions. An application of the equation to a broad
wave-packet shows that a small amendment to the usual relativistic
formalism can be helpful to demonstrate continuity with the non-relativistic case, 
although difficulties  remain when the proposed quantum state has a broad relativistic energy
distribution.        


\newpage
~\\{\bf 1.  Introduction}

The simplest approach to a relativistic description of a quantum
particle was proposed in 1926 by Klein and by Gordon~\cite{kg}, among
others~\cite{fok}; a historical survey has been given by
Kragh~\cite{kragh}.  The familiar relativistic equation $ E^2 = p^2c^2
+ (mc^2)^2$ is quantised by replacing the classical observables $E$
and $p^2$ by quantum observables $i\hbar\del/\del t$ and $-\hbar^2[
(\del/\del x)^2 +(\del/\del y)^2 +(\del/\del z)^2 ]$.  Thus, the
Klein-Gordon (KG) equation for the wave function $\psi$ is
\begin{equation}
-\hbar^2\frac{\del^2\psi}{\del t^2}  = \left(-\hbar^2c^2
\left[\frac{\del^2}{\del x^2} +\frac{\del^2}{\del y^2} +\frac{\del^2}{\del z^2}\right] +
m^2c^4 \right)\psi.
\label{eq1}
\end{equation}
Since the energy eigenstate of a free quantum particle is represented
by a plane wave of the form $\exp i(\mathbf{k.x}-\omega t)$, with
$\omega$ and $\mathbf{k}$ proportional to energy and momentum, 
the KG equation must hold unavoidably for such states of a spinless
relativistic quantum particle. Being linear in $\psi$, it  remains valid
for  linear combinations of these plane-wave eigenstates, just as does  
the similarly linear (non-relativistic) Schr\"odinger equation. Although it was
first proposed for massive scalar particles, it should apply to
other massive particles that are in a definite spin state.  

Soon after the equation's proposal, objections were raised that were not fully
answered, resulting in uncertainty that is still present about the
validity of relativistic quantum mechanics, and of the KG equation in
particular.  These objections are repeated without resolution in most textbooks on the
subject. However,  they are ignored in much of the research
literature and there has been no shortage of publications, far too many to 
refer to here, in which the KG equation is accepted, modified, applied in a 
large variety of contexts, or employed in various
interpretations of quantum theory. Particular examples would be the evaluation of CP 
violation properties of meson states and entanglement in particle systems.
The KG equation is certainly in use, and the 
situation can be puzzling to those encountering the subject. 

The purpose of the present paper is to examine to what
extent the claimed difficulties with the equation can be overcome,
with the aim of clarifying how the equation should be better understood
and where it can be confidently applied. We will concentrate on
 the more basic quantum mechanical issues and, in particular, we will  
not discuss field theory at any length.

~\\{\bf 2. Traditional difficulties with the Klein-Gordon equation}\\[1ex]
{\it (i)  The use of a second-order differential equation, and the negative-energy problem}

A problem raised early in the KG equation's history was that a
second-order differential equation, such as (\ref{eq1}), does not
normally allow the future values of its argument to be predicted
from a given set of starting values alone. Thus the wave-function
$\psi$ at an initial reference time would apparently be insufficient to allow 
the KG equation to determine the
particle's later state, and this was seen as unsatisfactory. This
criticism was used to cast doubt on the KG equation and to support a
need for a first-order equation, such as that of Dirac, but it can be
countered. If $\psi(\mathbf{x})$ is Fourier analysed at an initial 
time in terms of eigenstates of momentum $\mathbf k$, then its future
behaviour is well determined from this information alone, using the KG
equation for each component, provided that there is no ambiguity in
specifying $\omega$ for a given  $\mathbf k$.  This  
requirement becomes the key to understanding many of the issues raised
regarding the KG equation, which formally allows solutions with both
positive and negative values of $\omega$.

A contrast can be made with a scalar classical wave, for example a
sound wave.  If, at a given time, the amplitude of such a wave is specified as a function 
of position, this does not differentiate between components travelling in opposite
directions, and for a wave of this type a given $\mbfit{k}$ can have 
either sign of $\omega$.  For the KG equation, however, this mathematical
possibility is to be discounted on the grounds of unphysicality.  A
relativistic massive particle always has positive energy, which means that for
a given $\mathbf{k}$, only a positive value of $\omega$ can be
assigned. The negative-energy solutions must be rejected, and by doing this we
can ensure that the KG wave equation determines the particle's future
state unambiguously from its original state, as desired.

Dirac's first-order differential equation for spin-half fermions did not remove
the mathematical existence of negative-energy states, but
placed them in a new part of the spinor.  As is well known, his
interpretation of these states was as a ``sea'' of mostly
filled states in which unoccupied states or ``holes'' behave as
antiparticles, whose energy is positive compared to that of an
{\it occupied} negative-energy state.  This viewpoint led to the successful
expectation that the positron should exist, but it subsequently fell
out of favour although it was still supported by Pauli in
1955~\cite{pauli}.  However it will not work at all for bosons,
because the entire negative-energy sea could acquire unlimited numbers
of bosonic particles, and the vacuum would be unstable.  Thus the
negative-energy solutions remain problematic.

~\\{\it (ii)  The  question of ``incompleteness" }

The apparent incompleteness of a set of solutions that lacks
negative-energy contributions has been criticised \cite{sakurai}.  But
it is clear that the Fourier components in terms of $E$ or $\omega$
cannot in any case comprise a formally complete set for a massive real
particle, because $|E|$ cannot be less than $mc^2$; the energy
spectrum is obliged to have a cutoff at this value.  
The set of states in $\omega$ is in this sense incomplete, but we may instead 
rely on the states specified by
$\mathbf{k}$, which constitute a complete set of physical states of a 
positive-mass particle.  This seems perfectly satisfactory. 

To take a classical example, there is no problem in insisting that
Pythagoras' equation shall give only positive solutions for the
hypotenuse of a triangle, even though the square root of $x^2 + y^2$
can mathematically take negative values.  We state that these solutions are
geometrically invalid and ignore them.  An analogous approach can be applied to the KG
equation.  Indeed, it does not seem to disturb us that Einstein's
original equation $ E^2 = p^2c^2 + (mc^2)^2$ bears
negative-energy solutions; these are likewise ignored.
It might seem anomalous to retain them in the quantum context.

~\\{\it (iii)  Feynman's solution}

Feynman's well-known response was to say that negative-energy
solutions to (\ref{eq1}) (and also to the Dirac equation) corres\-pond to 
anti\-particle states with positive energy, but with the sign of $t$ reversed~\cite{feyn}.  
The antiparticle states, he supposed, are to be included within the same
set of solutions as the particle states.  A problem here, however, is
that two different treatments of time  are then implied within the same
equation and its solutions.  While this interpretation is generally accepted in connection
with Feynman diagrams, it is therefore very untidy
with regard to free particles and the KG equation, and it is unclear
what to do when a particle is its own antiparticle, as for example the
$\pi^0$ meson -- only one set of states is wanted in this case. 

A much clearer and more transparent approach to the KG
equation is to treat particles and their antiparticles as solutions to
separate but identical equations, with negative-energy solutions for free particles always 
disallowed. When appropriate, quantum super\-positions 
of particle and antiparticle states may then be constructed for a given 
system, as is common practice with neutral meson systems. On the other hand,
a superposition of positive-charge and negative-charge
states is never observed. The issue of combining particle and
antiparticle states thus needs to be settled on its own terms
in a given physical situation and is not a simple implication of the relativistic
quantum formalism.

Feynman's proposal has been applied principally in connection with 
Feynman diagrams in field theory, in which propagators -- that is, virtual
particles to which the KG equation does not apply -- are employed to
indicate the transfer of particle characteristics between different
vertices in a scattering process: ``streams of influence", we may say.
Amongst other things, propagators denote energy flow, which can be
positive or negative, since positive energy flowing out
of a vertex is the same as negative energy flowing into it.  Feynman's  
picture is therefore in essence dynamical, whereas quantum
mechanics (starting from de Broglie's equations) has a foundation in
free real-particle states.  But an understanding of
these  is necessary as a basis for Feynman's theory.

As pointed out above, the basic objection to the use 
of Feynman's idea with the KE equation is that it requires one 
symbol $t$ to denote  two different signs of physical time simultaneously in the 
same differential equation, and within its one set of solutions.  This connotational 
ambiguity cannot reasonably be accepted.

~\\{\it (iv)  The probability density question}

A third and more serious criticism of the KG equation concerns the particle's 
spatial probabiity density $\rho$ and current \mbfit{j}, as discussed in many
textbooks; that by  Desai~\cite{desai} has
been referenced here.  For both the non-relativistic and relativistic
cases these quantities should be related by the continuity equation
\begin{equation}
 \frac{\del\rho}{\del t} = -\NABLA\cdot\mbfit {j}. 
\label{eq4}
\end{equation}
For a non-relativistic
particle, $\rho = \psi^*\psi,$ while the current is given by
\begin{equation}
\mbfit{j} =\frac{\hbar}{2im}\left[\psi^*(\NABLA\psi) - \psi(\NABLA\psi^*)\right].
\label{eq2}
\end{equation}
The same equation for $\mbfit{j}$ is conventionally taken also in the
relativistic case, where $m$ is again the rest mass, but the
probability density is now written as
\begin{equation}
\rho = \frac{-\hbar}{2imc^2}\left[ \psi^*\frac{\del\psi}{\del t} -
\psi\frac{\del\psi^*}{\del t}\right].
\label{eq3}
\end{equation}
This equation is counterintuitive, however, although it satisfies (\ref{eq4}), 
and it generates suspicions
about the KG equation.  For a plane wave, it gives
$\rho = (E/mc^2)\psi^*\psi$, which is negative for negative $E$.  Once
more, negative energy values are deeply problematic, since a
probability density cannot be negative.  An alternative suggestion
from Pauli and Weisskopf was that we are no longer really discussing
particle density but charge density, which can take 
negative values~\cite{pandw}.  This proposal has been taken up by other
authors~\cite{dittrich}, and it evades some of the difficulties, but
it does not really address the original task of evaluating particle
density and current; the KG equation says nothing about electric
charge. Neutral particles remain problematic,\footnote{There have been
attempts to solve this issue.  Greiner~\cite{greiner}, following
Bjorken and Drell~\cite{bjd} and Feshbach and Villars~\cite{fv}, gives
an argument whose conclusion is that a particle such as a $\pi^0$ must
have a wave-function that is a real mathematical function. But this
would have to be a real sinusoidal wave, which cannot be correct,
since this is not an eigenstate of the momentum operator
$-i\hbar\NABLA$~!}  and an unsatisfactory discontinuity is introduced
between the relativistic and non-relativistic interpretations; after
all, a relativistic particle may be just a non-relativistic particle
viewed in a different reference frame.  In the end, this suggestion
would seem to introduce more problems than it solves.

~\\[1ex]{\bf 3.  The case of a broad wave-packet}

It is instructive to examine the situation in more detail by
considering a broad wave-packet moving in one dimension $x$ with mean
positive energy ${E} = \hbar{\omega}$ and group velocity
$v_g$.\footnote{This topic has been treated by Mosley~\cite{mosley}
with a different mathematical perspective from that offered here.}
To a good approximation its wave function can be represented as
\begin{equation} \psi = Af(x-\bar{x})e^{i(kx-\omega t)},
\label{eqg1}
\end{equation}
where $A$ is a normalisation constant.  The function $f$ describes the
envelope in $x$ of the wave-packet, whose mean value is $\bar{x}=x_0
+v_gt$ and is $x_0$ at time $t=0$.  The partial differentials of $f$
with respect to $x$ and $t$ are thus $f'$ and $-f'v_g$
respectively. To avoid the introduction of further momentum
components, $f$ is taken to be real.  For example, a Gaussian 
wave-packet with half-width $\sigma$ has
\begin{equation}
\psi = \left(\frac{1}{\sqrt{2\pi}\sigma}\right)^\half e^{-(x-\bar{x})^2/4\sigma^2}
e^{i(kx-\omega t)}.
\label{eqg2}
\end{equation}
As usual, $v_g$ equals $\del{\omega}/\del{k}$ and is given by $\hbar
k/m = p/m$ for the \schreq\ and $c^2k/\omega = c^2p/E$ for the KG
equation. This equals $p/\relval{m}$, where $\relval{m}=
\gamma m$ is the relativistic mass and $E=\relval{m}c^2$. (As
expected, the relativistic group velocity has an upper bound of $c$.)
We consider a sufficiently broad wave-packet such that the values of
energy and momentum have narrow widths and the group velocity is well
defined.  Broadening effects with time are neglected at this level of description.

For a packet given by (\ref{eqg1}), the non-relativistic case gives  
\begin{equation}
\frac{\del\rho}{\del t} 
= \psi^*\frac{\del\psi}{\del t} + \psi\frac{\del\psi^*}{\del t} 
= -2v_g|A|^2ff'
\label{eqg3}
\end{equation}
and
\begin{equation}
j = \frac{\hbar}{2im} \left[\psi^*\frac{\del \psi}{\del x} -
\psi\frac{\del \psi^*}{\del x}\right] = \frac{\hbar k}{m}|A|^2f^2.
\label{eqg4}
\end{equation}
As intuitively expected, $j$ is $v_g$ times the local value of
$\psi^*\psi$.  Its gradient in $x$ is $j' = 2v_g|A|^2ff'$, in required
agreement with (\ref{eqg3}) and (\ref{eq4}).
  
In the relativistic case, we start from the representation
for $\rho$  given by (\ref{eq3}).  The
wave-packet (\ref{eqg1}) now gives 
\begin{equation}
\rho = \frac{\hbar}{mc^2}A^*Af^2\omega = \frac{E}{mc^2}\psi^*\psi = \gamma\psi^*\psi,
\label{eqg5}
\end{equation} 
since the terms in $ff'$ cancel; $\gamma$ is the usual relativistic
factor for the particle and $\rho$ is clearly positive.  We obtain 
\begin{equation}
\frac{\del\rho}{\del t} = -2\frac{\hbar\omega}{mc^2} |A|^2ff'v_g = -2\gamma v_g|A|^2ff'.
\label{eqg6}
\end{equation}
Taking the expressions for $j$ unchanged from  (\ref{eq2}) and (\ref{eqg4}), we obtain the
required result
\begin{equation}
j' =  2\frac{\hbar k}{m}|A|^2ff' = 2\gamma v_g|A|^2ff' = -\del\rho/\del t.
\label{eqg7}
\end{equation}

The KG equation thus gives a well-defined account of a relativistic
particle whose wave function has the form of a broad wave-packet, but
the conventionally adopted expressions for probability density and current 
gradient now give results that contain a factor of $\gamma$.  Since (\ref{eq3}) 
and (\ref{eq2}) are purely conventional, and the
probability density must be normalised to unity anyway, 
it would be convenient to redefine them to remove this
factor.  In this way, equations (\ref{eq3}) and (\ref{eq2}) may be written as 
\begin{eqnarray}
\label{eqrel1}
\relval{\rho} &=& \gamma^{-1}\frac{-\hbar}{2imc^2}\left[ \psi^*\frac{\del\psi}{\del t} 
- \psi\frac{\del\psi^*}{\del t}\right],  \mathrm{~which~reduces~to~} \psi^*\psi,\\[1.75ex]
\mathrm{and~~~~} 
\relval{\mbfit{j}} &=& \gamma^{-1}\frac{\hbar}{2im}\left[\psi^*(\NABLA\psi) - \psi(\NABLA\psi^*)\right].
\label{eqrel2}
\end{eqnarray}
These amended equations are valid provided that the wave packet has an
acceptably well-defined value of $\gamma$. They represent the
probability density and current for a relativistic wave-packet more
naturally than (\ref{eq3}) and (\ref{eq2}), since the quantity $\psi^*\psi$ is non-negative and 
may now be given its usual interpretation, while (\ref{eq4}) continues
to hold.  The rest mass $m$ in (\ref{eq3}) and (\ref{eq2}) is
to be replaced by the relativistic mass $\relval{m} = m\gamma$,
which clearly expresses the continuity between the non-relativistic and
relativistic accounts.

~\\ {\bf 4. A more general case}

The example of the previous section does not give a complete answer to the probability 
density problem.  To investigate  further, we examine the 
case of a wave function  that is the sum of components with different energy.
Let
\begin{equation}
\psi = \sum_j{\psi_j},\  \mathrm{where}\:\; \psi_j = a_j e^{i(\mathbf{k_j.x} - \omega_jt)}.
\end{equation}
Thus $\del\psi_j/\del t = -i\omega_j\psi_j$.  The terms in the sum may have 
arbitrary positive energies $\hbar\omega_j$
and arbitrary amplitudes $a_j$, subject to unitarity.  Then (\ref{eq3}) gives
\begin{eqnarray}
\rho &=& \frac{\hbar}{2m}\sum_j{\omega_j\sum_k({\psi_k^*\psi_j + \psi_k\psi^*_j})} \nonumber \\
&=&   \frac{\hbar}{m}\sum_j{\omega_j\:{\Large (}|a_j|^2 + \sum_{k\neq j}|a_j a_k|\cos \phi_{ jk}{\Large )}},
\end{eqnarray}
where $ \phi_{ jk}$ is the phase difference between $\psi_j$ and $\psi_k$ at a given spacetime point 
$(\mathbf{x}, t)$. 
Even with the $\omega_j$ positive, $\rho$ is not necessarily positive in this situation.  
This is easily seen by considering the case of just two terms, which gives
\begin{equation} 
 \rho  = \frac{\hbar}{m}\left[\omega_1\left( |a_1|^2 + |a_1 a_2| \cos \phi_{1\,2}\right)
+ \omega_2\left( |a_2|^2 + |a_1 a_2| \cos \phi_{1\,2}\right)\right].
\end{equation}
 If the two amplitudes are equal in magnitude, then both terms are non-negative.
Otherwise, the term with smaller amplitude can become negative and does not have to be
cancelled everywhere by the other term if $\omega_1\neq\omega_2$.  
Thus $\rho$ can become negative, and this may be expected to  be true
in general for more than two components and 
for the case of a broad continuous distribution in $\omega$. 

~\\{\bf 5. Discussion}

We have identified the most major source of difficulty with the KG
equation as associated with solutions with negative energy; once these
are dismissed, many of the claimed problems disappear.  In practice, many applications 
of the theory ignore such states implicitly, but it is interesting to observe 
just how troublesome
they turn out to be.\footnote{Removing such states is also a feature of 
the Foldy-Wouthuysen treatment of the Dirac equation~\cite{greiner}.  Some aspects of the
present discussion can also be applied to the Dirac equation, but we
do not pursue this in detail here. Another approach, with doubtful
success, is to modify the KG equation so as to produce only positive-energy
solutions~\cite{kostin}.}

The question of apparently negative probability densities is more
complex. Most practical applications (including field treatments) concern 
particles in plane-wave states, although a more realistic approach
requires a wave-packet model.  Within the KG
equation, such states do not give a probability problem. 
(A plane wave has $f=1$ in eq.~(\ref{eqg1}).) 
 The state should have a well-defined central positive energy and a narrow energy
width. A non-relativistic particle wave function is also well described in this way when
viewed in a boosted reference frame, because its spread in
$\gamma$ remains small. This does not imply that the \schreq\
and its dynamics can be similarly transformed.  A proposal for a
more relativistic adaptation of the \schreq\ has been given by Grave
de Peralta et al.~\cite{gdp}.

Baym, unusually among textbook authors, says that the KG 
equation is ``quite useful", while repeating the problems that we 
have already discussed~\cite{baym}.  He shows that particle  
wave-packets that are narrowly confined in space are problematic.
This is no doubt true, and such particles also have a broad energy spectrum. 
However the states discussed in section 4.4 do not necessarily
have narrow spatial wave-packets, and so the difficulty appears to be more 
general than Baym indicates.

A particle state with a substantial spread of relativistic energy possesses
no proper rest frame and thus no non-relativistic counterpart. In the
absence of well defined energy, the redefinitions (\ref{eqrel1}) and
(\ref{eqrel2}) cannot be applied. In this case (\ref{eq3})
can give probability densities that are not always positive, although
the case of two contributing amplitudes of equal magnitude
remains well behaved.  Here, therefore, 
there apppear to be unresolved issues regarding the interpretation of
relativistic quantum theory;   the main problem is found with 
the probability density and current  rather than with the KG equation itself. 
A similar problem might also arise with regard
to a single-particle excited state of a quantum field, if it had poorly
defined relativistic energy. Such states are not commonly discussed, and 
this is a topic that invites further examination.

One simple suggestion would be that while the probability density
seems clearly measurable, corresponding to the position of a particle, 
its current may not always be a valid measurable variable.
Its interpretation, we have seen, is plausible in the case of a broad wave packet.  In
general, however, it does not correspond to a Hermitian quantum variable and so is not
a property of a particle that is measurable in the normal quantum scheme.  Perhaps,
then, it should not be considered as fundamentally important.

A limitation to any first-quantised relativistic theory is that dynamical
problems can be treated only approximately. It is perfectly possible
to solve the KG equation with a potential energy term included, for
example to evaluate a pionic atom~\cite{desai2, greinerpi}.  Here however, as
with the Dirac equation, a major issue is well known, namely that the
use of a simple potential does not give precise results in a relativistic
context.  Field theories are set up to avoid negative energies of real particles, 
and in the end provide a more comprehensive physical account, but we may still 
wish to explore the limits of the more basic quantum method.

~\\{\bf 6. Conclusions}

Special relativity relies in an essential way on the use of well-defined frames of reference.  
A quantum particle with a broad energy spread lacks a proper frame of reference, 
and so it may be no surprise if such states present interpretational difficulties.  
However if a free particle is in a state with positive energy that is 
sufficiently well-defined to provide a usable proper reference frame, the Klein-Gordon
equation gives an acceptable account of its basic quantum features, and the most frequent criticisms are 
overcome.  With a minor amendment to the standard notation, the
relativistic mass can now be used in the probability current equation,
giving a natural continuity between the non-relativistic and
relativistic treatments.  Despite the objections of many distinguished
practitioners of the subject, the negative-energy solutions to the
equation are not physically usable in describing single particles; 
real antiparticles should be treated
separately and  in a parallel way to the particles.  For many neutral meson
systems, the two classes of state can then be combined in an extended
quantum formulation.  

In the end, a relativistic equation of the Klein-Gordon type cannot be avoided for the 
description of  free spinless quantum particles and those with fixed-spin states. 
However, its application may be restricted
to states in which the particle has a proper relativistic rest frame.
With due allowance for this constraint, the Klein-Gordon equation 
is able to retain an important role in quantum mechanics. Its place is assured, provided that
limitations such as discussed above are kept in mind.

\vspace{2cm}
~\\
{\bf Acknowledgements.}  The author would like to thank the editors of 
Global Journal of Science Frontier Research for publishing this article free of charge
in GJSFR-A 22 (2022)  7(2)  and for help in the publication process.   
(https://globaljournals.org/journals/science-frontier-research/a-physics-space-science)
 

\end{document}